\documentclass{article}

\usepackage[utf8]{inputenc} 
\usepackage[T1]{fontenc}    
\usepackage{hyperref}       
\usepackage{url}            
\usepackage{booktabs}       
\usepackage{amsfonts}       
\usepackage{nicefrac}       
\usepackage{microtype}      

\usepackage[utf8]{inputenc} 
\usepackage[T1]{fontenc}    
\usepackage{hyperref}       
\usepackage{url}            
\usepackage{booktabs}       
\usepackage{amsfonts}       
\usepackage{nicefrac}       
\usepackage{microtype}      

\usepackage{mathrsfs,amsmath,amssymb,bm}
\usepackage{graphics,graphicx,epsfig,subfigure}

\def\B{{\mathcal{B}}}

\newtheorem{theorem}{Theorem}

\newtheorem{definition}{Definition}

\newtheorem{proposition}{Proposition}
\newtheorem{corollary}{Corollary}

\newcommand*{\QEDA}{\hfill\ensuremath{\blacksquare}}

\title{Convergence Analysis of Belief Propagation on Gaussian Graphical Models}

\author{
  Jian Du, Shaodan Ma, Yik-Chung Wu, Soummya Kar, and Jos\'{e} M. F. Moura
}

\begin{document}

\maketitle

\begin{abstract}
	Gaussian belief propagation (GBP) is a recursive computation method that is widely used in inference for computing marginal distributions efficiently.
	Depending on how the factorization of the underlying joint Gaussian distribution is performed, GBP may exhibit different convergence properties as different factorizations may lead to fundamentally different recursive update structures.
	In this paper,  we study the convergence of GBP derived from the factorization based on the distributed linear Gaussian model.
	The motivation is twofold. From the factorization
	viewpoint, i.e., by specifically employing a factorization based on the linear Gaussian model, in some cases, we are able to bypass difficulties that exist in other
	convergence analysis methods that use a different (Gaussian Markov random field) factorization.
	From the distributed inference viewpoint, the linear Gaussian model
	readily conforms to the physical network topology
	arising in large-scale networks, and, is practically useful.
	For the distributed linear Gaussian model, under mild assumptions,
	we show analytically three main results: the GBP message inverse variance converges exponentially fast to a unique positive  limit  for arbitrary nonnegative initialization;
	we  provide a necessary and sufficient convergence condition for the belief mean  to converge to the optimal value;
	and, when the underlying factor graph is given by the union of a forest and a single loop, we show that  GBP always converges.
\end{abstract}

\vspace{-1em}
\section{Introduction}
\vspace{-0.5em}
A multivariate Gaussian distribution $q(x)$ can be expressed in the information form as \cite{WalkSum1}
$$
q(x)\propto\exp\Big\{
-\frac{1}{2}x^TJx + h^Tx
\Big\},
$$
where $J$, termed as the information matrix, is a symmetric, positive definite matrix ($J\succ 0$) and $h$ is the  potential vector.
Conditional independence relationship among variables in $x$ can be viewed via an undirected graph also known as the
Gaussian Markov random field (GMRF).
To construct the GMRF, the underlying Gaussian distribution is first factorized as
\begin{equation}\label{MRF-eqn}
q\left( x\right)
\propto
\prod_{i\in \mathcal{V}} \psi_{i} \left( x_i \right)
\prod_{J_{i,j}\neq 0; i\leq j } \psi_{i,j} \left( x_i, {x}_j\right),
\end{equation}
where
\vspace{-0.5em}
\begin{equation}\label{pairwise}
\psi_{i} \left(x_i\right) = \exp\{-\frac{1}{2}J_{i,i}x^2_i + h_ix_i\}
\quad
\textrm{and} \quad
\psi_{i,j} \left(x_i, x_j\right) = \exp\{-x_iJ_{i,j}x_j\}.
\end{equation}

The associated GMRF  contains a set of vertices  corresponding to each random variable $x_i$,  and each vertex for $x_i$ is associated with the \textit{node potential function} $\psi_{i} \left(x_i\right)$.  If $J_{i,j}\neq 0$, there is an edge between $x_i$ and $x_j$, which is associated with the \textit{edge potential function} $\psi_{i} \left(x_i, x_j\right)$.
Fig. 1(a) is an example of the GMRF  corresponding to an instance of $J$ in (\ref{J}).

For inference problems, given a joint Gaussian distribution  $q(x)$, one  is often interested in computing the marginal distribution for each variable $x_i$ in $x$, which is equivalent to determining the mean vector
$\mu := \mathbb E\{x\} $ and diagonal elements of the covariance matrix $P := \mathbb E\{(x-\mu)(x-\mu)^T\} $ with
\begin{equation}\label{1}
\mu = J^{-1}h  \quad \textrm{and} \quad P = J^{-1}.
\end{equation}
However, directly inverting $J$ has  computational complexity  order
$\mathcal{O}\left(M^3\right)$ with $M$ being the dimension of $x$.
Moreover, in large-scale networks with random variables $x_i$ associated with different agents in the network, computing $J^{-1}$ at a fusion center may also suffers from large communication overhead, heavy computation burden, and be susceptible to central agent failure.
Dealing with highly distributed data  has been recognized by the U.S.  National Research Council as one of the big challenges for processing big data \cite{MassiveData}.
Therefore, distributed processing to infer  $\mu$  and diagonal elements of $P$ that
only requires local communication and local computation is important for problems arising in  distributed networks \cite{du2014TSP, du2016convergence}.

Gaussian belief propagation (GBP)
\cite{DiagnalDominant} provides an efficient  way for computing the marginal mean in (\ref{1}).
Although with great empirical success \cite{Murphy}, and as recognized  in different research areas, it is known that a major challenge that hinders GBP is the lack of convergence guarantees   in loopy networks.
Convergence of other forms of loopy BP are analyzed in \cite{chertkov2006loop, NIPS2016_6318,  gomez2007truncating,NIPS2012_4649,Ihler05, Martin13}, but  these analyses are not directly
applicable to  GBP.
All previous convergence analyses of GBP have  focused on GMRFs  \cite{DiagnalDominant,WalkSum1,minsum09}, where the the joint distribution is factorized  according to (\ref{MRF-eqn}).
Recently, a distributed convergence condition for both GMRF and Gaussian linear model is proposed in \cite{du2017verification}.
In  \cite{WalkSum1},
based on the fact that $(J^{-1})_{i,j}$ can be interpreted as the sum of the weights of all the walks from $j$ to $i$ on the corresponding GMRF, a sufficient convergence condition, given by the spectrum radius
$\rho(|I - J|)<1$, is obtained, commonly known as walk-summable property.
We emphasize two important points here. First, restricting to GMRFs, the recursive update structure of the GBP obtained on the basis of the usual GMRF factorization (\ref{MRF-eqn}) could be different from the recursive updates obtained using other types of factorizations such as those based on the distributed linear Gaussian model representation of the GMRF studied in this paper. Secondly, and importantly, there exist GMRF scenarios (see example below) in which the information matrix $J$ fails to satisfy the walk-summable property, although, a GBP update based on a different factorization of the GMRF (specifically, based on the distributed linear Gaussian model representation of the GMRF as studied in this paper) may be
obtained that is shown to yield convergence. Before proceeding further, consider the following GMRF in which the  information matrix $J$ is given by
\begin{equation}\label{J}
J=\left[
\begin{matrix}
\begin{smallmatrix}
1&\frac{1}{3\sqrt{2}}& \frac{1}{\sqrt{3}} & \frac{\sqrt{2}}{3}  \\
\frac{1}{3\sqrt{2}}&1&0&\frac{1}{3}	\\
\frac{1}{\sqrt{3}}&0&1&\frac{1}{\sqrt{6}}\\
\frac{\sqrt{2}}{3}&\frac{1}{3}&\frac{1}{\sqrt{6}}&1
\end{smallmatrix}
\end{matrix}
\right],
\end{equation}
which satisfies $J\succ 0$ and $J = J^T$.
Since $\rho(|I-J|)=1.0754$, which is non walk-summable  \cite{WalkSum1}, the convergence condition in \cite{WalkSum1} is inconclusive as to whether GBP  converges for this GMRF.
Rather than the GMRF factorization,
we study the GBP convergence of this example by employing a different  factorization method based  on the linear Gaussian model representation.
To this end, we rewrite
$J$  as
$J = A^TR^{-1}A + W^{-1}$, where
$
A=\left[
\begin{matrix}
\begin{smallmatrix}
\frac{2}{\sqrt{6}}& 0 & \frac{1}{\sqrt{2}} & \frac{1}{\sqrt{3}} \\
\frac{{1}}{\sqrt{6}}&\frac{{1}}{\sqrt{3}}&0&0	\\
0&\frac{{1}}{\sqrt{3}}&0&\frac{{1}}{\sqrt{3}}
\end{smallmatrix}
\end{matrix}
\right]$,
$
W=\left[
\begin{matrix}
\begin{smallmatrix}
6& 0 & 0 & 0 \\
0& 3&0&0	\\
0&0&2&0\\
0&0&0&3
\end{smallmatrix}
\end{matrix}
\right],
$
and $R = I$.
Note that  $J^{-1}=(A^TR^{-1}A + W^{-1})^{-1}$ is  the covariance matrix for the joint posterior distribution of the linear Gaussian model
\begin{equation}\label{linearC}
y = A {x} +  {z}
\end{equation}
with ${z}\sim \mathcal{N}\left( {z}| {0}, {R}\right)$ and $x\sim \mathcal{N}\left( {x}| {0}, W\right)$.
Consequently,
to perform the marginalization equivalently, GBP  is employed on an alternative
factorization of $q(x)$, given by
\vspace{-0.5em}
\begin{equation}\label{jointpost1}
q\left(  x\right)
\propto
p\left( {y}_1|   {x}_1,{x}_2,{x}_3  \right)
p\left( {y}_2|   {x}_1,{x}_2  \right)
p\left( {y}_3|   {x}_2,{x}_3  \right)
\prod_{n=1}^3
p\left(x_n\right).
\end{equation}
This factorization stems from the products of the local  likelihood functions and local prior distributions  associated with (\ref{linearC}).
This can be
expressed by a
factor graph,
where  every variable  $ {x}_i$ is represented by a circle (called a variable node)
and the probability distribution of a  variable or a group of  variables is represented by a square (called a factor node).
A variable node is connected to a factor node if the variable is an argument of that particular factor.
For example, Fig.~1(b) shows the factor graph representation for the GMRF in Fig.~1(a)
\footnote{Factor graphs represent the same Markov relationship among variables as  in the GMRFs with the same $q(x)$ \cite[Chapter~9]{mezard2009information}.}.
In Theorem \ref{ref}, we successfully show that, for a factor graph
that is the union of a forest and a single loop, as  in  Fig.~1(b), GBP always converges  to the exact $\mu_i$.
This is in sharp contrast to the fact that for the same joint distribution and with factorization based on the classical  GMRF  representation in (\ref{MRF-eqn}), existing conditions and analyses are inconclusive as to whether GBP will converge or not.

We note that the factor graph based representation of GMRFs and distributed linear Gaussian models as in (\ref{linearC})-(\ref{jointpost1})
arise
in a variety of areas including   image interpolation \cite{image-interp}, cooperative localization \cite{Win}, distributed beamforming \cite{A2}, distributed synchronization  \cite{JianClock, JianCFO, du2017proactive, du2013fully}, fast solver for system of linear equations \cite{A4},   factor analysis learning  \cite{A6},  sparse Bayesian learning \cite{A7}, and peer-to-peer rating in social networks \cite{A9}, in which it is of interest to compute the $\mu_{i}$ in a distributed fashion.
Recently, \cite{giscard2016exact} proposes an exact inference method to compute $\mu$ and $J^{-1}$ based on path sum on the GMRF with arbitrary topology. GBP, by contrast, only targets to compute $\mu$ and the diagonal elements of $J^{-1}$, the  parameters of the associated marginal  distributions.
Different from GBP that only requires local computation and communication, \cite{giscard2016exact} requires summing over all the simple paths and simple cycles on the graph, which requires  centralized processing for computing  all the simple paths and simple cycles as well as centralized scheduling.

\begin{figure}\label{SystemModel}
	\centering
	\mbox{\subfigure[]{\epsfig{figure=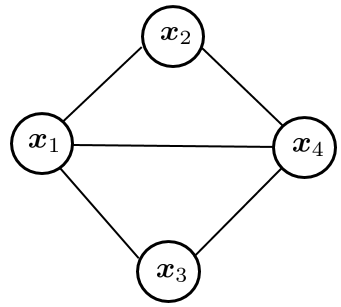,width=1.2in}}\label{Network} }
	\quad
	\mbox{ \subfigure[]{\epsfig{figure=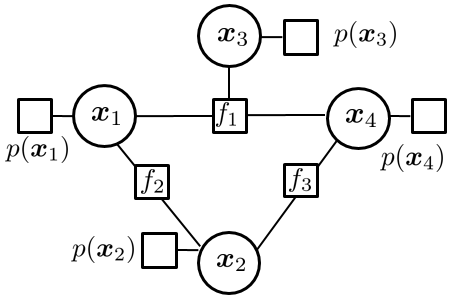,width=1.8in}}\label{MRF} }
	\caption{(a) The GMRF corresponding to $J$ in (\ref{J}) with the factorization following (\ref{MRF-eqn});
		(b) The factor graph  corresponding to $J$ in (\ref{J}) with the factorization following (\ref{jointpost}).
	}
\end{figure}

In this paper, we first  study analytically the convergence of
GBP for the linear Gaussian models.
The distributed  algorithm based on GBP involves  only  local computation and local communication, and thus, scales with the network size.
We  prove
convergence  of the
resulting GBP distributed algorithm.
Specifically, by establishing certain contractive properties of the dynamical system modeling the distributed inverse covariance updates under the Birkhoff metric, we show that,
with arbitrary nonnegative initial message  inverse variance,
the  belief variance  for each variable converges  at a geometric rate to a unique positive limit.
We
also demonstrate that
there is a better choice
for  initializing the message  inverse variance
than the commonly used  zero initial condition
to achieve faster convergence.
Further, we prove, under  the proposed necessary and sufficient convergence condition, the belief mean
converges to the optimal value.
In particular, we show that for  a   graph that is given by the union of a single loop and a forest, GBP always converges.

\section{Distributed Inference with GBP on  Linear Gaussian Model}
\label{headings}
Consider a general connected network
of $M$  agents, with $\mathcal{V}=\left\{1,\ldots, M\right\}$ denoting the set of agents.
Agent $n$ has an unknown random variable $x_n$ and makes a local linear observation
${y}_n = \sum_{i\in  n\cup\mathcal{I}\left(n\right)}
{A}_{n,i}{x}_i + {z}_n$,
where
$\mathcal{I}\left(n\right)$ denotes the set of  neighbors of agent $n$,
and
${A}_{n,i}$ is a known coefficient\footnote{This model also allows two neighboring agents to share a common observation \cite{du2017pairwise}, e.g., $y_i = y_j = G(x_i-x_j)+z_i$. Our analysis applies to this case.}.
The prior distribution for $x_i$ is Gaussian, i.e., ${x}_i\sim \mathcal{N}\left({x}_i|{0},{W}_{i}\right)$,
and ${z}_n$ is the additive noise with distribution ${z}_n\sim \mathcal{N}\left({z}_n|{0},{R}_n\right)$
\footnote{If the $y_n$ is noiseless, it would represent  linear equality constraints among the variables, i.e., ${y}_n = \sum_{i\in  n\cup\mathcal{I}\left(n\right)}
	{A}_{n,i}{x}_i$, which conflicts with $J>0$ that we assume. Thus, we  assume $R_n>0$ in this paper.}.
It is assumed that
$p\left({x}_i, {x}_j\right)=p\left({x}_i\right)p\left({x}_j\right)$
and
$p\left({z}_i,{z}_j\right)
=p\left({z}_i\right)p\left({z}_j\right)$ for $i\neq j$.
The goal is to learn ${x}_i$, based on ${y}_n$, $p\left({x}_i\right)$, and $p\left({z}_n\right)$.
The joint distribution  $p\left(  x\right)p\left( {y}|  x\right)$ can be written as the product of the  prior distribution and the local likelihood function as
\begin{equation}\label{jointpost}
p\left(  x\right)p\left( {y}|  x\right) =
\prod_{n\in \mathcal{V}}
p\left(x_n\right)
\prod_{n\in \mathcal{V}}
\underbrace{p\left( {y}_n| \left\{ {x}_i\right\}_{i\in  n\cup\mathcal{I}\left(n\right)} \right)}_{:= f_n }.
\end{equation}
We derive the Gaussian
BP algorithm   over the corresponding factor graph to learn $  x_n$ for all $n\in \mathcal V$. It
involves two types of messages:
one is the message
from a variable node $  x_j$ to its neighboring factor node $f_n$, defined as
\begin{equation} \label{BPv2f1}
m^{\left(\ell\right)}_{j \to f_n}\left(  x_j\right)
= p\left(  x_j\right)
\prod_{f_k\in \mathcal B(j)\setminus f_n}m^{\left(\ell-1\right)}_{f_k\to j}\left(  x_j\right),
\end{equation}
where $\mathcal B\left(j\right)$ denotes the set of neighbouring factor nodes  of $  x_j$,
and $m^{\left(\ell-1\right)}_{f_k\to j}\left(  x_j\right)$ is the   message  from $f_k$ to $  x_j$ at  $(\ell-1)$-iteration.
The second type of
message is from a factor node $f_n$ to a neighboring variable node $ {x}_i$, defined as
\begin{equation}\label{BPf2v1}
m^{\left(\ell\right)}_{f_n \to i}\left( {x}_i\right)
=  \int \cdots \int
f_n \times \!
\prod_{j\in\mathcal B\left(f_n\right)\setminus i} m^{ \left(\ell\right)}_{j \to f_n}\left(  x_j\right)
\,\mathrm{d}\left\{  x_j\right\}_{j\in\B\left(f_n\right)\setminus i},
\end{equation}
where $\mathcal B\left(f_n\right)$ denotes the set of neighboring variable nodes of $f_n$.
The process iterates between equations (\ref{BPv2f1}) and (\ref{BPf2v1}).
At each iteration $\ell$, the approximate marginal distribution, also referred to as belief, on $ {x}_i$ is computed locally at $ {x}_i$ as
\begin{equation} \label{BPbelief}
b_{\textrm{BP}}^{\left(\ell\right)}\left( {x}_i\right)
= p\left( {x}_i\right) \prod_{ f_n\in \mathcal B\left(i\right)} m^{\left(\ell\right)}_{ f_n \to i}\left( {x}_i\right).
\end{equation}

Let
the initial  messages  at each variable node and factor node  be
in Gaussian function  forms as
$m^{\left(0\right)}_{f_n \to i}\left( {x}_i\right)\propto
\exp
\big\{-\frac{1}{2}
|| {x}_i-  {v}^{\left(0\right)}_{f_n\to i}||^2
_{{J}^{\left(0\right)}_{f_n\to i}}
\big\}$
and $
m^{\left(0\right)}_{j \to f_n}\left(  x_j\right)\propto
\exp\big\{-\frac{1}{2}||  x_j- {v}^{\left(0\right)}_{j\to f_n}||^2_{  {J}^{\left(0\right)}_{j \to f_n}}\big\}.$
It is evident  that the general expression for the message from variable node $j$ to factor node $f_n$ is
$	m^{\left(\ell\right)}_{j \to f_n}\left(  x_j\right) \propto
\exp
\big\{-\frac{1}{2}
||  x_j- {v}^{\left(\ell\right)}_{j\to f_n}||^2
_{ {J}^{\left(\ell\right)}_{j\to f_n}}
\big\}$,
with
\begin{equation} \label{v2fV}
{J}^{\left(\ell\right)}_{j \to f_n}
=  {W}_j^{-1} +
\sum_{f_k\in\B\left(j\right)\setminus f_n}
{J}_{f_k\to j}^{\left(\ell-1\right)},
\quad
{v}^{\left(\ell\right)}_{j\to f_n}=
\left[{J}^{\left(\ell\right)}_{j\to f_n}\right]^{-1}
\sum_{f_k\in\B\left(j\right)\setminus f_n}
{J}_{f_k\to j}^{\left(\ell-1\right)}
{v}^{\left(\ell-1\right)}_{f_k\to j},
\end{equation}
where $ {J}_{f_k\to j}^{\left(\ell-1\right)}$ and $  {v}_{f_k\to j}^{\left(\ell-1\right)}$ are the message  inverse variance   and mean  received at variable node $j$ at the $\left(\ell-1\right)$-$\textrm{th}$ iteration, respectively.
Furthermore,
the message  from factor node  $f_n$ to variable node $i$ is given by
$	m^{\left(\ell\right)}_{f_n \to i}\left( {x}_i\right)\propto
\exp
\big\{-\frac{1}{2}
|| {x}_i- {v}^{\left(\ell\right)}_{f_n\to i}||^2
_{ {J}^{\left(\ell\right)}_{f_n\to i}}
\big\}$,
with
\begin{equation}\label{Cov}
{J}^{\left(\ell\right)}_{f_n\to i}
=
{A}_{n,i}^2
\Big[  {R}_n
+\!\!\!\!\!\!\!\!
\sum_{j\in\B\left(f_n\right)\setminus i}\!\!\!\! {A}_{n,j}^2 \left[{J}^{\left(\ell\right)}_{j\to f_n}\right]^{-1} \Big]^{-1}\!\!\!,
\quad
{v}^{\left(\ell\right)}_{f_n\to i}
=
{A}_{n,i}^{-1}
\Big( {y}_n-\!\!\!\!\!\!\!\!\sum_{j\in\B\left(f_n\right)\setminus i}
\!\!\!\! {A}_{n,j}
{v}^{\left(\ell\right)}_{j\to f_n}\Big).
\end{equation}

For this factor graph based approach, according to the message updating procedure  (\ref{v2fV}) and (\ref{Cov}),  message exchange is only needed between neighboring nodes.
For example, the messages transmitted from node $n$ to its neighboring node $i$ are   $m_{f_n\to i}^{\left(\ell\right)}\left( {x}_i\right)$ and $m_{n\to f_i}^{\left(\ell\right)}\left(  x_n\right)$.
Thus, the message passing scheme given in (\ref{BPv2f1}) and (\ref{BPf2v1})  conforms with the network topology.
Furthermore, if the messages $m_{j\to f_n}^{\left(\ell\right)}\left(  x_j\right)$
and $m_{f_n\to i}^{\left(\ell\right)}\left( {x}_i\right)$ exist for all $\ell$,
the messages  are Gaussian, therefore only the corresponding mean  and inverse of variance are needed to be exchanged.
Finally,  according to the definition of belief in (\ref{BPbelief}),
$b_{\textrm{BP}}^{\left(\ell\right)}\left( {x}_i\right)$ at  iteration  $\ell$ is computed as
$		{b}_{\textrm{BP}}^{\left(\ell\right)}\left( {x}_i\right)
=p\left( {x}_i\right)\prod_{f_n\in\mathcal B\left(i\right)} m_{f_n\to i}^{\left(\ell\right)}\left( {x}_i\right)
\propto  \mathcal{N}\big( {x}_i|
{\mu}_i^{\left(\ell\right)}, {P}_i^{\left(\ell\right)}\big)$,
where
\begin{equation}  \label{beliefcov}
{P}_i^{\left(\ell\right)} =
\Big[ {W}_i^{-1}
+\sum_{f_n\in\B\left(i\right)}
{J}_{f_n\to i}^{\left(\ell\right)}\Big]^{-1}
\quad
\textrm{and}\quad
{\mu}_i^{\left(\ell\right)}= {P}_i^{\left(\ell\right)}\Big[
\sum_{f_n\in\B\left(i\right)}
{J}_{f_n\to i}^{\left(\ell\right)} {v}^{\left(\ell\right)}_{f_n\to i}\Big].
\end{equation}
The iterative computation
starts by initializing
$	{J}_{f_k\to j}^{(0)}$ and
$	{v}_{f_k\to j}^{\left(0\right)}
$ (\ref{v2fV}) for all $k\in \mathcal V$ and $j\in \mathcal B(f_k)$;
it  terminates when message (\ref{v2fV}) and (\ref{Cov}) converges to a fixed value or the maximum number of iterations is reached.

\section{GBP Convergence Analysis}
\label{others}
\vspace{-0.5em}
A challenge with GBP  for large-scale networks is determining whether it  converges or not.
In particular, it
is generally known that, if the factor graph contains cycles, the GBP algorithm may
diverge.
Thus, determining  convergence conditions for the GBP algorithm is very important.
Sufficient conditions for the convergence of GBP in loopy graphs are available in \cite{DiagnalDominant, WalkSum1}.
However,  these conditions are derived based on the classical  GMRF based factorization of the joint distribution in the form of (\ref{pairwise}).
This  differs from the model considered in this paper, where the factor
$f_n$ follows (\ref{jointpost}), which leads to  intrinsically different recursive equations.
More specifically, the recursive equations
(\ref{v2fV}) and (\ref{Cov}) have different properties from recursive equations (7) and (8) in \cite{WalkSum1}.
Thus, the convergence results in \cite{DiagnalDominant, WalkSum1} cannot be applied to the GBP for the linear Gaussian model.

Due to the recursive  updating  of  $m_{j\to f_n}^{\left(\ell\right)}\left(  x_j\right)$ and $m_{f_n\to i}^{\left(\ell\right)}\left( {x}_i\right)$ in (\ref{v2fV}) and (\ref{Cov}), the message evolution can be simplified by combining these two messages into one.
By substituting
$  {J}^{\left(\ell\right)}_{j \to f_n}$ in (\ref{v2fV}) into   (\ref{Cov}), the updating of the message variance inverse can be written as
\vspace{-0.5em}
\begin{equation}\label{CovFunc}
{J}_{f_n\to i}^{\left(\ell\right)}
=\!
{A}_{n,i}^2 \bigg[ {R}_n\!
+\! \!\!\!\!\!\!\sum_{j\in\B\left(f_n\right)\setminus i}
\!\!\!\!\!\!\!\!
{A}^2_{n,j}
(
{W}_{j}^{-1} \!\!+\!\!\!\!\!\!\!
\sum_{f_k\in\B\left(j\right)\setminus f_n}
\!\!\!\!\!\!\!\!
{J}_{f_k\to j}^{\left(\ell-1\right)}
)^{-1}
\bigg]^{-1}\!\!\!\!
:=\!
\mathcal{F}_{n\to i}
\left(\left\{
{J}_{f_k\to j}^{\left(\ell-1\right)}\right\}_{\left(f_k, j\right)\in \mathcal{\widetilde{B}}\left(f_n, i\right)}
\right),
\end{equation}
where $\mathcal{\widetilde{B}}\left(f_n, i\right)=\left\{\left(f_k, j\right) | j \in \B\left(f_n\right)\setminus i,  f_k\in \B\left(j\right)\setminus f_n\right\}$.
Observing that ${J}_{f_n\to i}^{\left(\ell\right)}$ in (\ref{CovFunc}) is independent of $ {v}^{\left(\ell\right)}_{j\to f_n} $ and $ {v}^{\left(\ell\right)}_{f_n\to i}$ in (\ref{v2fV}) and (\ref{Cov}),
we can first focus on the convergence property of $ {J}_{f_n\to i}^{\left(\ell\right)}$ alone and then later on the convergence  of $ {v}^{\left(\ell\right)}_{f_n\to i}$.
Once we have  the convergence characterization of
${J}_{f_n\to i}^{\left(\ell\right)}$ and
$ {v}^{\left(\ell\right)}_{f_n\to i}$,  we will go back and investigate the convergence  of belief variances  and means in (\ref{beliefcov}).

\vspace{-0.5em}
\subsection{Convergence Analysis of Message Inverse Variance  }\label{Convariance}
To efficiently represent the updates of all message inverse variances, we {{introduce} the following definitions.
	Let
	${ {J}}^{\left(\ell-1\right)}
	\triangleq
	\texttt{Bdiag}
	\Big(\left\{ {J}_{f_n\to i}^{\left(\ell-1\right)}\right\}_{n\in \mathcal{V},i\in \B\left(f_n\right)}\Big)$ be
	a  diagonal  matrix with diagonal elements
	being the   message inverse variances  in the network at iteration $\ell-1$
	with index arranged in ascending order first on $n$ and then on $i$.
	Using the definition of $ {J}^{\left(\ell-1\right)}$, the term $\sum_{f_k \in \mathcal B\left(j\right) \backslash f_n}   {J}_{f_k\rightarrow j}^{\left(\ell-1\right)} $ in (\ref{CovFunc}) can be written as ${\Xi}_{n,j} J^{\left(\ell-1\right)} {\Xi}_{n,j}^T$, where ${\Xi}_{n,j}$  selects appropriate components from $ J^{\left(\ell-1\right)}$ to form the summation.
	Further, define $ {H}_{n,i}=\left[\left\{ {A}_{n,j} \right\}_{j\in B\left(f_n\right) \backslash i}\right]$,
	${\Psi}_{n,i}  = \texttt{Bdiag} \Big( \left\{ {W}_j ^{-1} \right\}_{j\in B\left(f_n\right) \backslash i} \Big)$ and
	$ {K}_{n,i}=\texttt{Bdiag} \left(\left\{ {\Xi}_{n,j} \right\}_{j\in B\left(f_n\right) \backslash i}\right) $, each with components  arranged in an ascending order on $j$.  Then (\ref{CovFunc}) can be written as
	\vspace{-0.5em}
	\begin{equation}\label{CovFunc3}
	{J}^{\left(\ell\right)}_{f_n\rightarrow i}= {A}_{n,i}^2\left\{ {R}_n+ {H}_{n,i}^2\left[ \Psi_{n,i} + {K}_{n,i}^2 \left( {I}_{|\mathcal{B}\left(f_n\right)|-1} \otimes  {J}^{\left(\ell-1\right)}\right)   \right]^{-1}    \right\} ^{-1}.
	\end{equation}
	Now,  define the function $\mathcal{F}\triangleq\left\{\mathcal{F}_{1\to k}, \ldots, \mathcal{F}_{n\to i}, \ldots, \mathcal{F}_{n \to M}\right\}$ by
	${ {J}}^{\left(\ell\right)} = \mathcal{F}\left({ {J}}^{\left(\ell-1\right)}\right) $.
	By stacking $ {J}_{f_n\to i}^{\left(\ell\right)}$ on the left side of
	(\ref{CovFunc3}) for all $n$ and $i$ as the  diagonal matrix $ {J}^{\left(\ell\right)}$, we obtain
	\begin{equation}\label{CovFunc5}
	{J}^{\left(\ell\right)}
	= {A}^T \Big \{ {\Omega}+ {H}\left[{\Psi} +  {K} \left({I}_\varphi \otimes {J}^{\left(\ell-1\right)}\right)  {K}^T \right]^{-1}  {H}^T \Big\} ^{-1} {A}, :=  \mathcal F\left( {J}^{\left(\ell-1\right)}\right),
	\end{equation}
	where $ {A}$, $ {H}$,
	${\Psi}$,  and $ {K}$ are  diagonal matrices with elements $ {A}_{n,i}$, $ {H}_{n,i}$, ${\Psi}_{n,i} $, and $ {K}_{n,i}$, respectively, arranged in ascending order, first on $n$ and then on $i$ (i.e., the same order as $ {J}^{\left(\ell\right)}_{f_n \rightarrow i}$ in $ {J}^{\left(\ell\right)}$).
	Furthermore,
	$\varphi={\sum _{n=1} ^M |\mathcal B\left(f_n\right)|\left(|\mathcal B\left(f_n\right)|-1\right)}$
	and
	${\Omega}$ is a block diagonal matrix with diagonal blocks $ {I} _{|B\left(f_n\right)|} \otimes  {R}_n$ with ascending order on $n$, where $\otimes$ denotes the matrix Kronecker product.
	We first present  some properties of the updating operator $\mathcal{F}\left(\cdot\right)$,
	which may be readily verified.
	
	\begin{proposition} \label{P_FUN}
		The updating operator $\mathcal{F}\left(\cdot\right)$ satisfies the following properties 	with
		respect to the partial order induced by the cone of positive semidefinite matrices.
	\end{proposition}

	\noindent P \ref{P_FUN}.1:
	$\mathcal{F}\left( {J}^{\left(\ell\right)}\right) \succeq \mathcal{F}\left( {J}^{\left(\ell-1\right)}\right)$, if $ {J}^{\left(\ell\right)} \succeq {J}^{\left(\ell-1\right)}\succeq {0}$.
	
	\noindent P \ref{P_FUN}.2: $\alpha\mathcal{F}\left( {J}^{\left(\ell\right)}\right) \succ  \mathcal{F}\left(\alpha {J}^{\left(\ell\right)}\right)$
	and
	$\mathcal{F}\left(\alpha^{-1} {J}^{\left(\ell\right)}\right) \succ  \alpha^{-1}\mathcal{F}\left( {J}^{\left(\ell\right)}\right)$, if $ {J}^{\left(\ell\right)} \succ  {0}$ and $\alpha>1$.
	
	\noindent P \ref{P_FUN}.3:
	Define
	$ {U}\triangleq  {A}^T  {\Omega}^{-1} {A}$
	and $ {L}\triangleq
	{A}^T \left[  {\Omega}+ {H}{\Psi}^{-1}  {H}^T \right] ^{-1} {A}$.
	With arbitrary $ {J}^{\left(0\right)}\succeq  {0}$,
	$\mathcal{F}\left( {J}^{\left(\ell\right)}\right)$ is bounded as
	$ {U} \succeq  \mathcal{F}\left( {J}^{\left(\ell\right)}\right)\succeq {L}\succ  {0}$ for $\ell\geq 1$.

	Note that we use $X \succeq 0$ in the paper to denote $X$ is a positive semi-definite matrix.
	Based on the above properties of $\mathcal{F}\left(\cdot\right)$, we  establish the convergence of $J$ \cite{du2017var}.

	\begin{theorem} \label{guarantee}
		The matrix sequence
		$\left\{ {J}^{\left(\ell\right)}\right\}_{l=0,1,\ldots}$ defined by (\ref{CovFunc5}) converges to a unique positive definite matrix
		for any initial  $ {J}^{\left(0\right)}\succeq  0$.
	\end{theorem}
	
	Proof Outline.
	The set $\left[ {L},  {U}\right] $  is a compact set.
	Further, according to P \ref{P_FUN}.3, for arbitrary
	$ {J}^{\left(0\right)}\succeq {0}$,  $\mathcal{F}$ maps $\left[ {L}, {U}\right] $  into itself starting from
	$\ell\geq 1$.
	Since $\left[ {L}, {U}\right]$ is also a convex set,
	the continuous function $\mathcal F$ maps a compact convex subset of the Banach space of positive definite matrices into itself.
	Therefore, the mapping $\mathcal F$ has a fixed point in $\left[ {L},  {U}\right]$ according to  Brouwer's Fixed-Point Theorem.
	The uniqueness of the fixed point can be proved by contradiction assuming there are more than one fixed point.  Leveraging  the properties of $\mathcal F(\cdot)$ in Proposition 1, we  can show that
	${J}^{\left(\ell\right)}_{l=0,1,\ldots}$ defined by (\ref{CovFunc5}) converges to a unique positive definite matrix
	for any initial covariance matrix.

	According to  Theorem \ref{guarantee}, $ J^{\left(\ell\right)}_{f_n\to i}$ converges  if all initial message inverse variances are nonnegative, i.e.,  ${J}^{\left(0\right)}_{f_n\to i}\geq  {0}$
	for all $i \in \mathcal V$ and $f_n \in \mathcal B\left(i\right)$.
	{Notice that, for the GMRF model, the message inverse variance does not necessarily converge for all initial non-negative values.}
	Moreover, due to the computation of ${J}^{\left(\ell\right)}_{f_n\to i}$  being independent of the local observations $ {y}_n$,
	as long as the network topology does not change, the converged value  $ {J}^{\ast}_{f_n\to i}$ can be precomputed offline and stored at each node, and there is no need to re-compute ${J}^{\ast}_{f_n\to i}$ even if $ {y}_n$ varies.

	Another fundamental question is how fast the convergence is, and this is the focus of the discussion below.
	Since the convergence of  a dynamical system is often studied with respect to the part metric \cite{PartBook},
	in the following, we
	start by introducing the part metric.
	
	\begin{definition}\label{mydef}
		Part (Birkhoff) Metric \cite{PartBook}:
		For arbitrary square matrices $ {X}$ and $ {Y}$ with the same dimension,
		if there exists
		$\alpha\geq 1$ such that $\alpha {X} \succeq  {Y} \succeq \alpha^{-1}  {X} $,
		$ {X} $ and $ {Y}$
		are called the parts,
		and
		$ \mathrm{d} \left( {X}, {Y}\right)\triangleq
		\inf \left\{\log \alpha: \alpha {X} \succeq  {Y}\succeq \alpha^{-1}  {X}, \alpha \geq 1\right\}$
		defines a metric  called the part metric.
	\end{definition}

	Next, we will show that
	$\left\{ {J}^{\left(\ell\right)}\right\}_{l=1,..}$
	converges at a geometric rate  with respect to the part metric in $\mathcal J$ that is constructed as
	\begin{equation}\label{partset}
	\mathcal J =\left\{ {J}^{\left(\ell\right)}|   {U} \succeq  {J}^{\left(\ell\right)} \succeq {J}^{\ast}+ \epsilon  {I}\right\}
	\cup
	\left\{ {J}^{\left(\ell\right)}|
	{J}^{\ast}- \epsilon  {I} \succeq  {J}^{\left(\ell\right)} \succeq {L}\right\},
	\end{equation}
	where  $\epsilon>0 $ is a scalar and can be arbitrarily small.
	\begin{theorem}\label{RateCov}
		Let the initial message inverse variance   nonnegative, i.e., ${J}^{\left(0\right)}_{f_n\to i}\geq  {0}$. Then
		the sequence $\left\{ {J}^{\left(\ell\right)}\right\}_{\ell=0,1,\ldots}$ converges at a geometric rate  with respect to the part metric in $\mathcal J$.
	\end{theorem}
	\noindent {Proof}:
	{Consider two matrices $ {J}^{\left(\ell\right)} \in \mathcal C$, and $ {J}^{\ast} \not\in \mathcal C$.
		According to Definition \ref{mydef},
		we have
		$ \mathrm{d} \left( {J}^{\left(\ell\right)}, {J}^{\ast}\right)\triangleq
		\inf \left\{\log\alpha: \alpha {J}^{\left(\ell\right)} \succeq {J}^{\ast}\succeq \alpha^{-1} {J}^{\left(\ell\right)}\right\}$.
		Since $\mathrm{d} \left( {J}^{\left(\ell\right)}, {J}^{\ast}\right)$ is the smallest number satisfying $\alpha  {J}^{\left(\ell\right)} \succeq  {J}^{\ast} \succeq \alpha^{-1} {J}^{\left(\ell\right)}$, this is equivalent to
		$	\exp\left\{\mathrm{d} \left( {J}^{\left(\ell\right)}, {J}^{\ast}\right)\right\} {J}^{\left(\ell\right)}
		\succeq
		{J}^{\ast}
		\succeq
		\exp\left\{-\mathrm{d} \left( {J}^{\left(\ell\right)}, {J}^{\ast}\right)\right\}
		{J}^{\left(\ell\right)}.$
		Applying P \ref{P_FUN}.1 to this equation, we have
		$\mathcal F\left(\exp\left\{\mathrm{d} \left( {J}^{\left(\ell\right)}, {J}^{\ast}\right)\right\}
		{J}^{\left(\ell\right)}\right)
		\succeq
		\mathcal F\left( {J}^{\ast}\right)
		\succeq
		\mathcal F\left(\exp\left\{-\mathrm{d} \left( {J}^{\left(\ell\right)}, {J}^{\ast}\right)\right\}
		{J}^{\left(\ell\right)}\right).
		$
		Then applying P~\ref{P_FUN}.2 and
		considering  that   $\exp\left\{\mathrm{d} \left( {J}^{\left(\ell\right)}, {J}^{\ast}\right)\right\}>1$ and
		$\exp\left\{-\mathrm{d} \left( {J}^{\left(\ell\right)}, {J}^{\ast}\right)\right\}<1$, we obtain
		$	\exp\left\{\mathrm{d} \left( {J}^{\left(\ell\right)}, {J}^{\ast}\right)\right\}
		\mathcal F\left( {J}^{\left(\ell\right)}\right)
		\succ
		\mathcal F\left( {J}^{\ast}\right)
		\succ
		\exp\left\{-\mathrm{d} \left( {J}^{\left(\ell\right)}, {J}^{\ast}\right)\right\}
		\mathcal F\left( {J}^{\left(\ell\right)}\right)$.
		Notice that, for arbitrary positive definite matrices $ {X}$ and $ {Y}$, if $ {X}-k {Y}\succ  {0}$, then, by definition, we have $ {x}^T {X} {x}
		-k {x}^T {Y} {x}
		> {0}$ with $x\neq 0$.
		Then, there must exist $o>0$ that is small enough such that
		$ {x}^T {X} {x}
		-\left(k+o\right)
		{x}^T {Y} {x}
		> {0}$
		or equivalently
		$ {X}
		\succ \left(k+o\right)
		{Y}$.
		Thus, as $\exp{\left(\cdot\right)}$ is a continuous function, there must exist some $\triangle\mathrm{d}>0$ such that
		$
		\exp\left\{-\triangle\mathrm{d}+\mathrm{d} \left( {J}^{\left(\ell\right)}, {J}^{\ast}\right)\right\}
		\mathcal F\left( {J}^{\left(\ell\right)}\right)
		\succ
		\mathcal F\left( {J}^{\ast}\right)
		\succ
		\exp\left\{\triangle\mathrm{d}-
		\mathrm{d} \left( {J}^{\left(\ell\right)}, {J}^{\ast}\right)\right\}
		\mathcal F\left( {J}^{\left(\ell\right)}\right).\nonumber
		$
		Now, using  the definition of the part metric,  the above equation is equivalent to
		$	-\triangle\mathrm{d}+\mathrm{d}
		\left( {J}^{\left(\ell\right)}, {J}^{\ast}\right)
		\geq
		\mathrm{d} \left(\mathcal F\left( {J}^{\left(\ell\right)}\right), \mathcal F\left( {J}^{\ast}\right)\right)$.
		Hence, we obtain
		$\mathrm{d} \left(\mathcal F\left( {J}^{\left(\ell\right)}\right), \mathcal F\left( {C}^{\ast}\right)\right)
		<
		\mathrm{d}
		\left( {J}^{\left(\ell\right)}, {J}^{\ast}\right)$.
	}
	This result holds for any $ {J}^{\left(\ell\right)} \in \mathcal C$,
	$\mathrm{d} \left(\mathcal{F}\left( {J}^{\left(\ell\right)}\right), \mathcal{F}\left( {J}^{\ast}\right)\right) <
	c
	\mathrm{d} \left( {J}^{\left(\ell\right)}, {J}^{\ast}\right) $,
	where $c=\sup_{ {J}^{(\ell)}\in \mathcal{J}} \frac{\mathrm{d}\left(\mathcal{F}\left( {J}^{(\ell)}\right) , \mathcal{F}\left( {J}^{\ast}\right)\right) }{\mathrm{d} \left( {J}^{(\ell)}, {J}^{\ast}\right)}<1$.
	Consequently,
	we have
	\begin{equation}\label{rate}
	\mathrm{d} \left( {J}^{\left(\ell\right)}, {J}^{\ast}\right) <
	c^{\ell}
	\mathrm{d} \left( {J}^{\left(0\right)}, {J}^{\ast}\right).
	\end{equation}
	Thus the sequence $\left\{ {J}^{\left(\ell\right)}\right\}_{\ell=1,\ldots}$ converges
	at a geometric rate  with respect to the part metric. \QEDA

	The physical meaning of Theorem \ref{RateCov} is that the distance between $ {J}^{\left(\ell\right)}$ and $ {J}^{\ast}$ decreases exponentially before $ {J}^{\left(\ell\right)}$ enters the neighborhood of $ {J}^{\ast}$, which can be chosen to be arbitrarily small.
	Note that, for GBP based on the usual GMRF based representation and factorization, the message inverse variance convergence rate is still unknown .
	
	Moreover, we can further study  how to choose the initial value
	$ {J}^{\left(0\right)}$ so that $ {J}^{\left(\ell\right)}$ converges faster.
	Since $\mathcal F$ is a monotonic function,
	with
	$ {0} \preceq {J}^{\left(0\right)}\preceq  {L}$,
	$ {J}^{\left(\ell\right)}$ is a monotonically increasing sequence, and
	$ {J}^{\left(\ell\right)}$ converges most rapidly with
	$ {J}^{\left(0\right)}=  {L}$.
	Likewise, with
	$ {J}^{\left(0\right)}\succeq {U}$,
	$ {J}^{\left(\ell\right)}$ is a monotonically decreasing  sequence, and
	$ {J}^{\left(\ell\right)}$ converges most rapidly
	with
	$ {J}^{\left(0\right)}=
	{U}$.	
	It is common practice in the GBP literature to set  the initial message inverse variance  to  $ {0}$, i.e.,
	${J}^{\left(0\right)}_{f_n\to i}= {0}$
	\cite{DiagnalDominant, WalkSum1}.
	The above analysis, by contrast, reveals that there is a better
	choice  to guarantee faster convergence.
	\vspace{-0.5em} 	
	\subsection{Convergence Analysis of  Message Mean}
	According to Theorems \ref{guarantee} and \ref{RateCov},
	as long as we choose ${J}_{f_k\to j}^{\left(0\right)}\geq {0}$
	for all $j\in \mathcal V$ and $f_k\in \mathcal B\left(j\right)$,
	${J}_{f_k\to j}^{\left(\ell\right)}$
	converges at a geometric rate to a unique positive value
	${J}_{f_k\to j}^{\ast}$.
	Furthermore, according to (\ref{v2fV}),
	$
	{J}^{\left(\ell\right)}_{j \to f_n}$ also converges to a positive value once
	${J}_{f_k\to j}^{\left(\ell\right)}$ converges, and the converged value is denoted by
	$
	{J}^{\ast}_{j \to f_n}$.
	Then for arbitrary initial value
	${v}^{\left(0\right)}_{f_k\to j}$, the evolution of
	${v}^{\left(\ell\right)}_{j\to f_n}$ in
	(\ref{v2fV}) can be written in terms of the converged message inverse variances, which is
	\begin{equation}\label{v2fm36}
	{v}^{\left(\ell\right)}_{j\to f_n}=
	\left[{J}^{\ast}_{j\to f_n}\right]^{-1}
	\sum_{f_k\in\B\left(j\right)\setminus f_n}
	{J}_{f_k\to j}^{\ast}
	{v}^{\left(\ell-1\right)}_{f_k\to j}.
	\end{equation}
	\vspace{-0.5em}
	Using (\ref{Cov}), and
	replacing indices $j$, $i$, $n$ with $z$, $j$, $k$ respectively,
	${v}^{\left(\ell-1\right)}_{f_k\to j}$  is given by
	\begin{equation}\label{f2vmm37}
	\begin{split}
	{v}^{\left(\ell-1\right)}_{f_k\to j}
	=
	\left[{J}_{f_k\to j}^{\ast}\right]^{-1}
	{A}_{k,j}^T
	\Big[{{R}_k
		+
		\sum_{z\in\B\left(f_k\right)\setminus j} {A}_{k,z}^2
		[{J}^{\ast}_{z\to f_k}]^{-1}
	} \Big]^{-1}
	\Big({y}_k-\sum_{z\in\B\left(f_k\right)\setminus j} {A}_{k,z}
	{v}^{\left(\ell-1\right)}_{z\to f_k}\Big).
	\end{split}
	\end{equation}
	Let ${M}_{k,j} = {R}_k
	+
	\sum_{z\in\B\left(f_k\right)\setminus j} {A}_{k,z}^2
	[{J}^{\ast}_{z\to f_k}]^{-1}$.
	Substituting (\ref{f2vmm37}) into
	(\ref{v2fm36}), we have
	$ {v}^{\left(\ell\right)}_{j\to f_n}=
	{b}_{j \to f_n}-
	\sum_{f_k\in\B\left(j\right)\setminus f_n}
	\sum_{z\in\B\left(f_k\right)\setminus j}
	[{J}^{\ast}_{j\to f_n}]^{-1}
	{A}_{k,j}^T {M}_{k,j}^{-1}
	{A}_{k,z}
	{v}^{\left(\ell-1\right)}_{z\to f_k}$,
	where $ {b}_{j\to f_n}=
	[{J}^{\ast}_{j\to f_n}]^{-1}
	\sum_{f_k\in\B\left(j\right)\setminus f_n}
	{A}_{k,j} {M}_{k,j}^{-1}
	{y}_k $.
	The above equation can be further written in compact form as
	\begin{equation}\label{v2fm4}
	{v}^{\left(\ell\right)}_{j\to f_n}=
	{b}_{j \to f_n} -
	{Q}_{j \to f_n}
	{v}^{\left(\ell-1\right)},
	\end{equation}
	with the column vector $ {v}^{\left(\ell-1\right)}$
	containing  $ {v}^{\left(\ell-1\right)}_{z\to f_k}$ for all $z\in \mathcal V$ and $f_k\in \mathcal B\left(z\right)$ as subvector
	with ascending index first on $z$ and then on $k$.
	The matrix   $ {Q}_{j \to f_n}$
	is a row  vector with components
	$ {C}^{\ast}_{j\to f_n}
	{A}_{k,j}^T {M}_{k,j}^{-1}
	{A}_{k,z}$
	if $f_k\in\B\left(j\right)\setminus f_n$ and ${z\in\B\left(f_k\right)\setminus j}$, and $ {0}$ otherwise.
	Let $ {Q}$ be the  matrix that stacks  $ {Q}_{j \to f_n}$ first on $j$ and then on $ n$,
	and  $ {b}$ be the vector  containing
	$ {b}_{j \to f_n}$ with the same stacking order as $ {Q}_{j \to f_n}$. We have
	\begin{equation}\label{meanvectorupdate}
	{v}^{\left(\ell\right)} =
	- {Q}  {v}^{\left(\ell-1\right)} + {b}.
	\end{equation}
	It is well known that, for arbitrary initial value $ {v}^{\left(0\right)}$, $ {v}^{\left(\ell\right)}$ converges
	if and only if the spectral radius $\rho\left( {Q}\right)<1$.
	As $ {v}^{\left(\ell\right)}$ depends on the convergence of  $ {J}^{\left(\ell\right)}$, we have the following result.
	\begin{theorem} \label{meanvector}
		The vector sequence
		$\left\{ {v}^{\left(\ell\right)}\right\}_{l=0,1,\ldots}$ defined by (\ref{meanvectorupdate}) converges to a unique value
		for any initial value $\left\{ {v}^{\left(0\right)}\right\}$ and initial $ {J}^{\left(0\right)}\succeq  0$ if and only if $\rho\left( {Q}\right)<1$.
	\end{theorem}
	
	Though the above analysis provides a necessary and sufficient condition for the convergence of $ {v}^{\left(\ell\right)}_{j\to f_n}$, the condition $\rho\left( {Q}\right)<1$ needs to be  checked before using the distributed algorithm with equations (\ref{v2fV}) and (\ref{Cov}).
	In the sequel, we will show that
	$\rho \left( {Q}\right) <1$ for a
	union of a forest and a single loop
	factor graph; thus GBP converges  in such a topology.
	Although \cite{Weiss2000}
	shows the convergence of BP on the GMRF with a single loop, the analysis method cannot be applied here since the factorization of the joint distribution in   (\ref{jointpost})  is different from the pairwise model in \cite{Weiss2000}.
	
	\begin{theorem}\label{ref}
		For a connected factor graph that is the union of a forest and a single loop,
		with arbitrary  $ {J}^{\left(0\right)}_{f_n\rightarrow i}\geq  {0}$ for all $i\in \mathcal V$ and $f_n\in \mathcal B\left(i\right)$, the message inverse variance ${J}^{\left(\ell\right)}_{f_n\rightarrow i}$ and mean
		$ {v}^{\left(\ell\right)}_{f_n\rightarrow i}$ is guaranteed to
		converge to their corresponding unique points.
	\end{theorem}
	Proof Outline.
	Note that, for a single loop factor graph that is the union of a forest and a single loop, there are two kinds of nodes.
	One  is the factors/variables in the loop; the other  is the factors/variables on the chains/trees but outside the loop.
	Then message from a variable node to a neighboring factor node on the graph can be categorized into three groups:
	1) messages on a tree/chain passing towards the   loop;
	2) messages on a tree/chain passing away from the loop; and
	3) messages in the loop.
	According to (\ref{BPv2f1}), computation of  the messages in the first group does not depend on messages in the loop and is thus convergence guaranteed.
	Also,  from the definition of message computation in
	(\ref{BPv2f1}),
	if messages in the third group converge, the second group messages  should also converge.
	Therefore, we  focus on showing the convergence of messages in the third group.
	By observing  that in the single loop case $QQ^T$ is a diagonal matrix and  showing that the diagonal elements are all smaller than $1$, we can show that $\rho(Q)<1$. Therefore, for a factor graph that is the union of a forest and a single loop, GBP algorithm converges.

	\vspace{-0.5em}
	\subsection{Convergence Analysis of Belief Variance and Mean }\label{C}
	According to the first equation in (\ref{beliefcov}), as the computation of the belief variance $ {P}_i^{\left(\ell\right)}$ depends on  $
	{J}_{f_n\to i}^{\left(\ell\right)}$,
	using
	Theorems \ref{guarantee}
	and   \ref{RateCov}, we have the following important
	corollary that reveals the convergence and uniqueness property of $ {P}_i^{\left(\ell\right)}$.
	\vspace{-0.5em}
	\begin{corollary} \label{C-converge-iff2}
		With arbitrary initial message inverse variance ${J}^{\left(0\right)}_{f_n\to i} \geq  {0}$ for all $i\in \mathcal V$ and $f_n \in \mathcal B\left(i\right)$,
		the belief variance $ {P}_i^{\left(\ell\right)}$ converges to a unique positive value
		at a geometric rate with respect to the part metric in $\mathcal J$, where $\mathcal J$ is defined in (\ref{partset}).
	\end{corollary}
	\vspace{-0.5em}
	Proof outline. Since $P_i^{(\ell)}$ is a nonlinear function of ${J}^{\left(\ell\right)}_{f_n\to i}$ according to (\ref{beliefcov}), convergence analysis of $P_i^{(\ell)}$ is not trivial.
	In \cite{du2016convergence}, we  provide
	some  additional properties of the part metric  to facilitate the proof of the convergence rate of $P_i^{(\ell)}$.

	On the other hand, the computation of the belief mean ${\mu}_i^{\left(\ell\right)}$ depends on the belief variance $ {P}_i^{\left(\ell\right)}$ and on the message mean $ {v}^{\left(\ell\right)}_{f_n\to i}$ as shown in (\ref{beliefcov}).
	Thus, under the same condition as  in Theorem \ref{meanvector}, ${\mu}^{\left(\ell\right)}_i$ is convergence guaranteed.
	Moreover,  it is shown in \cite[Appendix]{MRFtoFG} that, for GBP over a factor graph,  the converged value of the belief mean equals its optimal value.
	Together with the convergence guaranteed topology
	shown in Theorem \ref{ref}, we have the following theorem.
	\vspace{-0.5em}
	\begin{theorem} \label{mean-if2}
		With  arbitrary  $ {J}_{f_n\to i}^{\left(0\right)}\geq   0$
		and $  v^{\left(0\right)}_{f_n\to i}$ for all $i\in \mathcal V$ and $f_n \in \mathcal B\left(i\right)$,
		the mean   ${\mu}_i^{\left(\ell\right)}$ in  (\ref{beliefcov}) converges
		to the optimal value ${  \mu}_i$ in (\ref{1})
		if and only if $\rho\left( {Q}\right)<1$,
		where $ {Q}$ is defined in (\ref{meanvectorupdate}).  Furthermore, the condition  $\rho\left( {Q}\right)<1$ is guaranteed to hold when the
		factor graph  is the union of a forest and a single loop.
	\end{theorem}
	
	\vspace{-1em}
	\section{Conclusions}
	\vspace{-1em}
	In this paper, we have studied the convergence properties of Gaussian belief propagation (GBP) that applies to  joint distributions
	factorized on the basis of linear Gaussian
	models.
	This kind of factorization leads to  GBP recursive computation structure that is different from existing work, in which GBP is directly applied to Gaussian Markov random fields (GMRFs) and factorizations based on the GMRFs.
	As demonstrated with an example, there are
	GMRF scenarios in which existing GBP convergence conditions based on the classical GMRF based factorization fail to apply.
	But when the underlying GMRF is represented as a linear Gaussian model and factorized according to the latter,
	the resulting GBP  converges as shown in this paper.
	For GBP applied to  joint distributions
	factorized based on linear Gaussian
	models,
	we have shown analytically that,
	with arbitrary nonnegative initial message inverse variance,
	the belief variance for each variable converges  at a geometric rate to a unique positive value.
	We have
	demonstrated that to guarantee faster convergence
	there is a better choice
	for the initial value
	than the commonly used all-zero initial condition.
	Moreover, we have presented  a necessary and sufficient  condition for convergence under which the belief mean
	converges to its optimal  value.
	We have also shown that  GBP always converges
	when the underlying graph is the union of a forest and a single loop.
	
	\small

\end{document}